\documentclass[iop]{emulateapj}
\usepackage[breaklinks,colorlinks,citecolor=blue,linkcolor=magenta,urlcolor=blue]{hyperref}
\newcommand*{\mysub}[2]{\ensuremath{#1_{\mathrm{#2}}}}

\newcommand*{\Omegam}{\mysub{\Omega}{m}}

\newcommand*{\Omegal}{\mysub{\Omega}{\Lambda}}

\newcommand*{\LCDM}{$\Lambda$CDM }

\newcommand*{\Msun}{\ensuremath{\, M_{\odot}}}

\newcommand*{\ltsim}{\ {\raise-.75ex\hbox{$\buildrel<\over\sim$}}\ }
\newcommand*{\gtsim}{\ {\raise-.75ex\hbox{$\buildrel>\over\sim$}}\ }
\newcommand*{\proptosim}{\ {\raise-.75ex\hbox{$\buildrel\propto\over\sim$}}\ }

\newcommand*{\secref}{Section}

\newcommand*{\Chandra}{{\it Chandra }}
\newcommand*{\Planck}{{\it Planck }}
\newcommand*{\Newfirm}{{\it NEWFIRM }}
\newcommand*{\Sqiid}{{\it SQIID }}
\newcommand{\Herschel}{{\it Herschel }}
\newcommand{\Xmm}{XMM-{\it Newton} }

\shorttitle{Three candidate clusters around radio-loud sources}
\shortauthors{Franck et al.}

\begin{document}
\title{Three candidate clusters around high redshift radio-loud sources: MG1 J04426+0202, 3C 068.2, MS 1426.9+1052}

%% Use \author, \affil, and the \and command to format
%% author and affiliation information.

\author{J.R. Franck\altaffilmark{1},
	 S.S. McGaugh\altaffilmark{1}, 
	 J.M. Schombert\altaffilmark{2}}

 \altaffiltext{1}{Case Western Reserve University, 10900 Euclid Ave., Cleveland, OH 44106}
 \altaffiltext{2}{University of Oregon, 1585 E. 13th Ave., Eugene, OR 97403}

\begin{abstract}
We present near-infrared observations of the environments around three radio-loud sources (MG1 J0442+0202, 3C 068.2, and MS 1426.9+1052) at redshifts $z=1.10,1.57$, and $1.83$ (respectively), that are surrounded by near-infrared galaxy overdensities. Overdensities with respect to field counts were found to be significant up to 19$\sigma$, with twelve times the expected number of galaxies within the inner regions of the densest proto-cluster. Color-magnitude relations are constructed in $K_s$, $J-K_s$, with each candidate cluster exhibiting a feature consistent with the beginnings of a red sequence. Galaxy models based on the redshift of the radio source are used to compare expected color-magnitude relations for a given formation epoch with the observed red sequence of each candidate, and are found to be consistent with an old ($z_f > 5$) formation epoch for a few bright, red galaxies on the red sequence. 
\end{abstract}


\section{Introduction} \label{sec:intro}

The largest overdensities in the early universe are observed today as galaxy clusters. These initial density perturbations can be used as probes over a range of epochs for the cosmological parameters that characterize our universe \citep{co94,vik09}. Cluster observables (number density, mass) as a function of redshift provide constraints for $\Omegal$ and $\Omegam$. The mass and formation timescale of high redshift clusters compared with simulations of large-scale structure is a powerful test of \LCDM \citep{mor11}. Recent discoveries are challenging the boundaries at how rapidly large structures can form ($z\sim 2$ from \citealt{gob13}; $z\sim 3.8$ from \citealt{lee14}; $z\sim 4$ from \citealt{ven02}), and how massive they can become \cite[e.g. `El Gordo' $M> 10^{15} \, \Msun$ at $z\sim 0.9$ in][]{men12}. Identification of the epoch of cluster formation is an important test of cosmology and the dark energy equation of state.

Previously these high redshift structures were identified via galaxy overdensities in near-infrared (NIR) photometric surveys \citep{ste03,pap08,eis08,pap10}. Clusters are also associated around radio-loud sources \citep{hal01} out to redshifts $z\sim 4$ \citep{ven02}. Utilizing the intra-cluster medium (ICM) as an identifier, diffuse X-ray emission has been useful in increasing the number of known high redshift clusters \citep{gob11,wil13}, while the Sunyaev-Zel'Dovich (SZ) effect is becoming increasingly powerful with the success of the South Pole Telescope (\citealt{kei11}), Atacama Cosmology Telescope (\citealt{sie13}), the \Planck satellite \citep{ade14a} , and with the Combined Array for Research in Millimeter-wave Astronomy (CARMA\footnote{\url{http://www.mmarray.org}}; \citealt{muc07,cul10,man14}). 

We present three cluster candidates at $z>1$, two of which have not been identified previously. High-redshift radio-loud sources were targeted for NIR photometry to search for a galaxy overdensity in the surrounding region, which has proven to be a robust identifier of structure \citep{ste03,haa09,gal10}. \secref~\ref{sec:obs} describes the NIR observations and photometry, while the overdensity associated with each cluster candidate is discussed in \secref~\ref{sec:over}. The photometry of each candidate cluster is presented in \secref~\ref{sec:cand} and analyzed in \secref~\ref{sec:jim}, with a summary presented in \secref~\ref{sec:sum}.

The work presented here assumes a \LCDM concordance cosmology, with $\Omegal = 0.7$, a matter density of $\Omegam = 0.3$, and $H_0 = 70$ km s$^{-1}$  Mpc$^{-1}$.

\section{Observations} \label{sec:obs}

\begin{deluxetable*}{c c c c c}
%\tabletypesize{\scriptsize}
\tablewidth{0pt}
\tablecolumns{6}
\tablecaption{Observations Log}
\tablehead{
\colhead{Date} & \colhead{Field} & \colhead{Instrument} & \colhead{Bands} & \colhead{Typical Seeing} \\
} 
\startdata
7-9 January 2006 & CL J0442+0202 & \Sqiid & $J,H,K_s,L$ & 0.5"-0.8"  \\ 
7-9 January 2006 & 3C 068.2	 & \Sqiid & $J,H,K_s,L$ & 0.5"-0.8" \\ 
7-9 April 2007 & MS 1426.9+1052 & \Sqiid & $J,H,K_s,L$ & 1.1"\\ 
12-13 October 2008 & 3C 068.2 & \Newfirm & $J,H,K_s$ & 1.1"-1.3"
\enddata
\tablecomments{Observations Summary for the three selected candidate clusters.   \label{tab:log}}
\end{deluxetable*}

Three candidate clusters were observed with Kitt Peak National Observatory's Mayall
4-m telescope equipped with the \Sqiid Camera in the NIR $J,H,K_s$ and $L$ passbands
between January 2006 to April 2007. A summary of observations can be found in
Table~\ref{tab:log}. Each passband is projected onto a $440 \times 460$ pixel InSb
detector (covering $172'' \times 179''$ in RA/DEC).  Individual frames were co-added
and corrected for cosmic rays with IRAF\footnote{IRAF is distributed by the National
Optical Astronomy Observatories, which are operated by the Association of
Universities for Research in Astronomy, Inc., under cooperative agreement with the
National Science Foundation.} \citep{iraf86,iraf93} {\it IMCOMBINE} routine.

The candidate cluster around 3C 068.2 was also imaged with the \Newfirm camera on the
4-m in October 2008.  \Newfirm is a wide-field camera that spans $28' \times 28'$,
and images were taken in $J,H,$ and $K_s$ filters. The cluster fell entirely within
one quadrant of the \Newfirm camera, so we devised the ``four-shooter'' observational
mode in which the telescope is shifted from quadrant to quadrant to maximize time on
target while simultaneously observing sky. These frames were processed by the
\Newfirm Quick-Reduce Pipeline.  The galaxy photometry between the \Newfirm and \Sqiid
images were within the photometric errors, increasing our confidence in the
calibration and detection procedures.

Sky subtraction was accomplished by chopping on and off field while dithering between
frames.  The resulting sky flats displayed variations of only 0.02\% based on sky box
determination of the frame sky value.  Hysteresis from a bright star in the $K_s$
frames for cluster MS 1426.9+1052 produced some large scale features (visible under
high contrast, see Figure 5).   However, these variations are less than 0.04\% the
background and do not effect the galaxy photometry routines which used local sky.
Both $J$ and $K_s$ frames were used for galaxy identification and each object was
confirmed visually in both frames before inclusion in our analysis.

Calibration of the stacked images utilized the 2MASS Point Source Catalog (PSC) for
stars within the individual frames in the $J$ and $K_s$ filters. With the wide field
of view (FOV) of {\it NEWFIRM}, a large sample of stellar sources were used for
calibration (474 matches for $J$ and 577 stars for $K_s$), providing photometric
solutions with intrinsic scatter of $\sigma_{J} = 0.05$ mag and $\sigma_{Ks} = 0.09$
mag. A photometric solution for the \Sqiid instrument was found in a similar way, but
the smaller FOV of each image provided fewer 2MASS calibration sources. CL J0442+0202
and 3C 068.2 were imaged on the same run in January 2006, while MS 1426.9+1052 was
observed in April 2007. Unfortunately, no usable 2MASS stars were in the latter
field, so the photometric solutions for MS 1426.9+1052 were derived using 2MASS stars
within other science frames taken during the April 2007 run.

Source Extractor ($SExtractor$; \citealt{sex}) was used to create an initial list of
sources from both $J$ and $K_s$ frames. Galaxies and stars were distinguished using
the stellar index parameter.  A few of the faintest objects were below the detection
threshold in one filter (typically $J$), and were confirmed visually, in both
filters.  A few objects were visible in the $K_s$ frames, but rejected as they were
below the threshold or not visible in the $J$ frame preventing a color determination.
The final set of objects are marked in Figures 1, 3 and 5, all clearly visible under
high contrast.  Overdensity estimates are made using a conservative limiting
magnitude in the $K_s$ frame, well below the detection threshold.

After the photometric solution was applied to the candidate cluster galaxies, a
color-magnitude relation (CMR) was constructed using $4''$ diameter apertures. The
aperture size was determined by building a curve of growth for the samples and
examining the source light profiles with {\it IMEXAM}.  The $4''$ diameter apertures
were found to be a suitable balance of light collected versus noise levels.

\section{Cluster-Candidate Galaxy Overdensities} \label{sec:over}

\begin{deluxetable*}{c c c c c c c}
%\tabletypesize{\scriptsize}
\tablewidth{0pt}
\tablecolumns{7}
\tablecaption{Summary}
\tablehead{
\colhead{Field} & \colhead{Redshift} & \colhead{RA} & \colhead{DEC} & \colhead{Galaxies Measured} & \colhead{Overdensity} & \colhead{Peak Surface Density}\\
\colhead{} & \colhead{} & \colhead{} & \colhead{} & \colhead{} & \colhead{} & \colhead{$arcmin^{-2}$}
} 
\startdata
CL J0442+0202 & $z = 1.10$ &  4:42:23.7 & +02:02:20 & 48 & 19.2$\sigma$ &  42.4	\\ 
3C 068.2	 & $z=1.57$ & 2:34:23.8 & +31:34:17 & 31 & 14.4$\sigma$ &  35.4 \\ 
MS 1426.9+1052 & $z=1.83$ & 14:29:24.1 & +10:39:15 & 33	& 12.5$\sigma$ & 32.4  
\enddata
\tablecomments{Summary Table for the three selected candidate clusters. All radio-loud sources have significant galaxy overdensities surrounding them with respect to the NIR field counts provided by \citet{hal98}. The Surface Density (in galaxies $\mathrm{arcmin}^{-2}$) is measured within the same area as the overdensity for sources $K_s < 20.5$.
\label{tab:Sum}}
\end{deluxetable*}

The most basic technique of identifying clusters is by finding overdensities of galaxies with respect to field counts. The lack of a large overdensity, conversely, does not imply that no cluster exists, as cluster richness and the system's mass were not found to be strongly correlated \citep{ade14b}, although other recent results have suggested the opposite \citep{2014ApJ...783...80R}. Following the example of \citet{hal98} and \citet{hal01}, counts of galaxies brighter than $K_s = 20.5$ surrounding radio-loud quasars consistently show overdensities relative to field counts ($13.7 \pm 1.5 \, arcmin^{-2}$). Listed in Table~\ref{tab:Sum} is each radio source, its position and redshift, and number of galaxies identified within $\sim 1'$ of the radio source.

%For MS1426, the most distant source is 80.5" distant. 13.7*[pi*80.5^2 /3600(" per ')] = 77.47 galaxies!

For each cluster candidate, a peak overdensity was determined by counting galaxies with $K_s < 20.5$ around the radio source, and comparing this to the expected field counts. The difference between the counted and expected (in units of $\sigma= 1.5$ counts $\mathrm{arcmin}^{-2}$) are presented in Table~\ref{tab:Sum} as the galaxy overdensity. The peak surface density for each candidate is computed as the number of galaxies divided by the same area as the peak overdensity, written in units of galaxies per square arcminute. 

The lowest redshift source in this sample was found to have the largest overdensity of 19.2$\sigma$ above the mean value, followed by 3C 068.2 with a 14.4$\sigma$ overdensity. The greatest redshift candidate cluster (at $z=1.83$) around MS 1426.9+1052 had the smallest overdensity of 12.6$\sigma$. This can be attributed to the off-center position of MS 1426.9+1052 in the \Sqiid image (see Fig \ref{fig:MSi}) effectively limiting the potential number of galaxies, coupled with the smaller amount of time the cluster candidate has had to dynamically assemble ($\sim$2 Gyrs between $z=1.11\rightarrow1.83$).

The expected number of field galaxies that reside within the region associated with each overdensity can be estimated utilizing the counts of \citet{hal98}. Within the areas associated with the peak overdensities (e.g. $>10\sigma$), the estimated number of galaxies un-associated with the cluster is $\sim 4/13$, $\sim 4/10$, and $\sim 4/10$ for CL J0442+0202, 3C 068.2, and MS 1426.9+1052, respectively. As each cluster candidate's overdensity is contained within approximately the same area on the sky, they have approximately the same number of expected field galaxies. 

It is important to note that galaxies which exist outside of the largest overdensity can still be associated with the cluster. For instance, out of the 6 spectroscopically confirmed members within CL J0442+0202, Galaxy 4 in Tab ~\ref{tab:CLt} is in an area outside an even modest overdensity ($<3\sigma$). Within the cluster discovered by \citet{pap10} at $z=1.62$, 8 of the 11 spectroscopically confirmed members would exist outside of a $\sim1\sigma$ overdense region if these galaxies were inserted into our data. This suggests that at these redshifts, clusters are still in the process of assembly with galaxies infalling.

By computing the expected number of galaxies for a projected area in square arcseconds around the radio source, an overdensity factor can also be computed. The inner regions show a factor of $\sim$ 11.6, 6.9, and 6.1 times the expected number of galaxies near CL J0442+0202, 3C 068.2, and MS 1426.9+1052, respectively. These overdensities indicate that the three regions are candidates for clusters in the process of assembly.

\section{Cluster Candidates} \label{sec:cand}

\subsection{CL J0442+0202}

%\clearpage
\begin{deluxetable}{c c c c c c}
%\tabletypesize{\scriptsize}
\tablewidth{0pt}
\tablecolumns{6}
\tablecaption{CL J0442 Sources}
\tablehead{
\colhead{Galaxy} & \colhead{$\Delta$RA } & \colhead{$\Delta$DEC} & \colhead{$\Delta$r } & \colhead{$K_s$} & \colhead{$J-K_s$}\\
\colhead{} & \colhead{('')} & \colhead{ ('')} & \colhead{('')} & \colhead{(mag)} & \colhead{(mag)} }
\startdata
1	&0.000	&	0.000	&	0.000	&	17.06	$\pm$	0.19	&	1.73	$\pm$	0.21	\\	
2	&27.571	&	-37.814	&	46.798	&	18.11	$\pm$	0.20	&	1.52	$\pm$	0.22	\\	
3\tablenotemark{b}	&42.527	&	-17.161	&	45.859	&	17.95	$\pm$	0.20	&	2.15	$\pm$	0.23	\\	
4\tablenotemark{a}	&28.336	&	-15.598	&	32.345	&	18.67	$\pm$	0.20	&	1.73	$\pm$	0.24	\\	
5	&4.610	&	-13.686	&	14.441	&	20.70	$\pm$	0.49	&	0.79	$\pm$	0.58	\\	
6\tablenotemark{c}	&-9.483	&	-11.521	&	14.922	&	18.12	$\pm$	0.20	&	1.30	$\pm$	0.22	\\
7\tablenotemark{b}	&-23.240	&	-9.315	&	25.037	&	17.59	$\pm$	0.19	&	1.73	$\pm$	0.21	\\	
8	&6.097	&	-7.555	&	9.709	&	17.75	$\pm$	0.19	&	0.88	$\pm$	0.21	\\	
9	&23.292	&	-3.630	&	23.574	&	18.80	$\pm$	0.21	&	1.25	$\pm$	0.23	\\	
10	&-0.907	&	4.254	&	4.350	&	19.54	$\pm$	0.26	&	1.74	$\pm$	0.37	\\	
11\tablenotemark{a}	&-7.788	&	5.418	&	9.487	&	18.11	$\pm$	0.20	&	1.63	$\pm$	0.22	\\
12\tablenotemark{c}	&-2.979	&	11.825	&	12.194	&	19.17	$\pm$	0.22	&	0.86	$\pm$	0.25	\\
13	&13.204	&	14.311	&	19.472	&	18.50	$\pm$	0.20	&	1.63	$\pm$	0.23	\\	
14\tablenotemark{a}	&3.693	&	14.121	&	14.596	&	19.20	$\pm$	0.22	&	1.59	$\pm$	0.28	\\	
15	&-28.050	&	29.989	&	41.063	&	19.07	$\pm$	0.21	&	1.59	$\pm$	0.26	\\	
16	&-21.010	&	-46.483	&	51.011	&	19.51	$\pm$	0.25	&	0.72	$\pm$	0.28	\\	
17	&15.502	&	34.860	&	38.151	&	19.30	$\pm$	0.23	&	1.97	$\pm$	0.33	\\	
18\tablenotemark{b}	&17.970	&	40.284	&	44.111	&	18.38	$\pm$	0.20	&	1.62	$\pm$	0.23	\\
19	&33.668	&	28.809	&	44.311	&	19.74	$\pm$	0.26	&	1.93	$\pm$	0.40	\\	
20	&-12.425	&	-8.076	&	14.819	&	19.53	$\pm$	0.23	&	1.95	$\pm$	0.38	\\	
21	&-8.792	&	0.011	&	8.792	&	20.16	$\pm$	0.41	&	1.39	$\pm$	0.58	\\	
22	&-31.158	&	-2.055	&	31.226	&	20.04	$\pm$	0.30	&	0.73	$\pm$	0.34	\\	
23	&-33.491	&	1.069	&	33.508	&	19.62	$\pm$	0.25	&	1.17	$\pm$	0.30	\\	
24	&-11.603	&	-34.214	&	36.128	&	19.25	$\pm$	0.22	&	1.86	$\pm$	0.31	\\	
25	&-32.993	&	-37.985	&	50.313	&	18.05	$\pm$	0.20	&	0.72	$\pm$	0.21	\\	
26	&-15.997	&	-49.336	&	51.865	&	19.34	$\pm$	0.23	&	1.38	$\pm$	0.28	\\	
27	&66.671	&	-35.440	&	75.505	&	18.99	$\pm$	0.21	&	1.08	$\pm$	0.24	\\	
28\tablenotemark{a}	&4.659	&	23.257	&	23.719	&	19.12	$\pm$	0.22	&	1.82	$\pm$	0.29	\\	
29	&-6.299	&	26.279	&	27.024	&	19.10	$\pm$	0.21	&	1.89	$\pm$	0.28	\\	
30	&20.863	&	29.580	&	36.197	&	19.92	$\pm$	0.30	&	0.91	$\pm$	0.34	\\	
31	&36.724	&	19.865	&	41.752	&	20.09	$\pm$	0.31	&	1.47	$\pm$	0.43	\\	
32	&11.888	&	-43.960	&	45.539	&	19.24	$\pm$	0.22	&	1.72	$\pm$	0.31	\\	
33	&56.662	&	20.766	&	60.347	&	19.67	$\pm$	0.27	&	0.99	$\pm$	0.31	\\	
34	&40.150	&	49.944	&	64.082	&	20.14	$\pm$	0.34	&	0.85	$\pm$	0.40	\\	
35	&43.020	&	55.760	&	70.426	&	19.22	$\pm$	0.23	&	1.13	$\pm$	0.26	\\	
36	&23.452	&	56.852	&	61.500	&	19.18	$\pm$	0.22	&	1.45	$\pm$	0.27	\\	
37	&51.838	&	50.321	&	72.245	&	18.54	$\pm$	0.20	&	2.58	$\pm$	0.30	\\	
38	&25.158	&	-0.324	&	25.160	&	19.53	$\pm$	0.24	&	1.59	$\pm$	0.32	\\	
39\tablenotemark{c}	&7.734	&	52.856	&	53.419	&	18.55	$\pm$	0.20	&	1.71	$\pm$	0.24	\\	
40	&14.390	&	38.199	&	40.820	&	20.47	$\pm$	0.38	&	0.87	$\pm$	0.44	\\	
41	&68.910	&	39.558	&	79.457	&	19.06	$\pm$	0.22	&	2.02	$\pm$	0.33	\\	
42	&42.107	&	48.183	&	63.989	&	19.64	$\pm$	0.25	&	1.49	$\pm$	0.33	\\	
43	&23.571	&	26.444	&	35.424	&	20.29	$\pm$	0.34	&	0.84	$\pm$	0.40	\\	
44	&32.678	&	10.572	&	34.346	&	19.87	$\pm$	0.27	&	2.35	$\pm$	0.60	\\	
45	&-29.289	&	13.467	&	32.236	&	19.49	$\pm$	0.24	&	2.20	$\pm$	0.41	\\	
46	&-27.731	&	-15.182	&	31.615	&	20.02	$\pm$	0.30	&	1.90	$\pm$	0.52	\\	
47	&25.050	&	-16.000	&	29.724	&	19.86	$\pm$	0.27	&	1.31	$\pm$	0.35	\\	
48\tablenotemark{a}	&-2.0272	&	-2.370	&	3.118	&	18.43	$\pm$	0.20	&	1.76	$\pm$	0.24	
\enddata
\tablenotetext{a}{Spectroscopically confirmed cluster members \citep{ste03}.}
\tablenotetext{b}{Non-members in the field \citep{ste03}.}
\tablenotetext{c}{X-ray point sources found with \Chandra that are not cluster members \citep{ste03}.}
\tablecomments{The central radio source MG1 J04426+0202 (Galaxy 1) and the sources around it within the \Sqiid field. Galaxy 1 is located
at RA=4:42:23.7 DEC=+02:02:20 with separations ($\Delta$RA,$\Delta$DEC) computed with respect to it. \label{tab:CLt}}
\end{deluxetable}

\begin{figure*}
\centering
\includegraphics[scale=0.5]{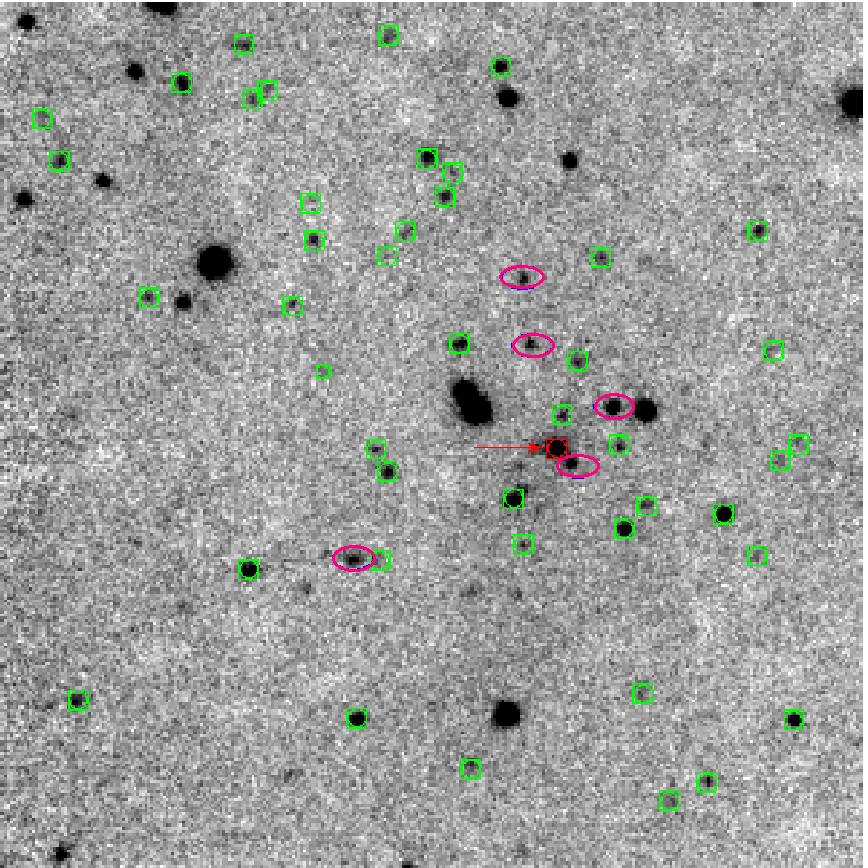}
\caption{The \Sqiid $K_s$ image of CL J0442+0202 that spans $120'' \times 120''$. The radio source MG1 J04426+0202 is indicated by a red arrow and square, while spectroscopically confirmed cluster members \citep{ste03} are inside magenta ellipses. The green boxes are the remainder of the sources measured photometrically.}
\label{fig:CLi}
\end{figure*}

%\clearpage
\begin{figure*}
\centering
\includegraphics[scale=0.5]{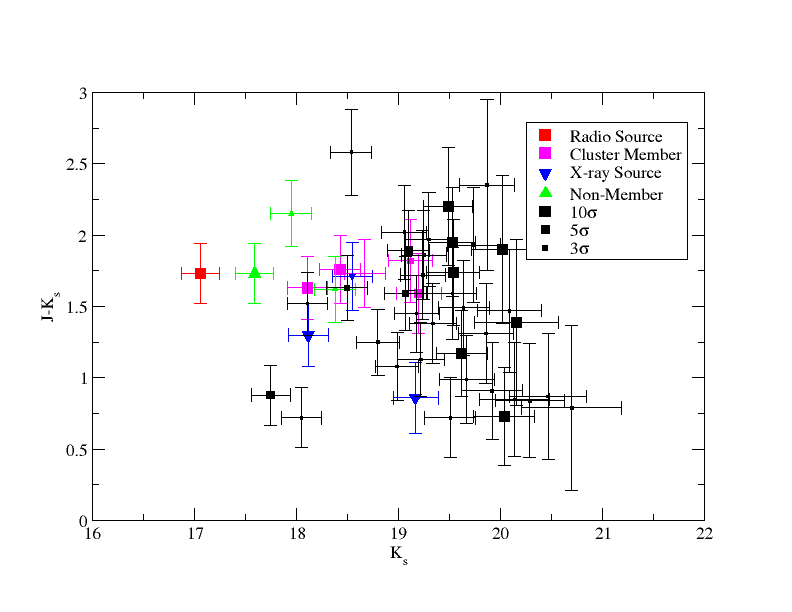}
\caption{The radio-loud source (MG1 J04426+0202) at the center of CL J0442+0202 is shown by a red square as the brightest cluster galaxy (BCG), and is also a part of the red sequence ($J-K_s =1.73$) in this color-magnitude diagram. The next brightest galaxy along the red sequence (a green triangle at $K_s=17.59$, $J-K_s=1.73$), is a spectroscopically confirmed interloper. Magenta squares are confirmed members \citep{ste03}, while foreground/background X-ray emitting AGN are represented by blue downward triangles. Galaxies that are found within the region corresponding to at least a 10$\sigma$ galaxy overdensity have the largest symbols, while the smallest points correspond to galaxies located in a region with an overdensity $<3\sigma$. Within the largest overdensity region, approximately 4 sources are expected to be unassociated with the cluster.}
\label{fig:CLf}
\end{figure*}

\citet{ste03} discovered a NIR galaxy overdensity surrounding the radio source MG1 J04426+0202 at a redshift of $z=1.11$, and did follow-up observations spectroscopically and in the X-ray with \Chandra. Five other galaxies were spectroscopically confirmed to be cluster members, and a number of X-ray point sources were resolved. However, \citet{ste03} were not able to detect diffuse X-ray emission from the ICM, only from the central radio source and other point sources. The NIR photometry of the cluster was measured in the $I$-band and Lick $K'$ to a limiting magnitude of $K' < 20$. 

We present \Sqiid photometry to a deeper limiting magnitude of $K_s < 20.5$. A comparison of 10 galaxies with $K' < 19.1$ magnitudes in the sample of \citet{ste03} with ours reveals a systematic offset of $K' - K_s = 0.32 \pm 0.16$ mag. This provides a basic check of the photometric solution for the \Sqiid camera, as well as a comparison to published work. The offset could be at least partially attributed to the blueward shift of the $K'$ filter. The uncertainty in the offset approximately mirrors the uncertainty within our own photometry, while the uncertainty of the $K'$ photometry from \citet{ste03} is unknown.

Within the field of view (shown in Fig~\ref{fig:CLi}), we present the photometry of 48 galaxies in Table~\ref{tab:CLt} and the resulting color-magnitude relation in Fig~\ref{fig:CLf}. Spectroscopically confirmed members \citep{ste03} are consistent with being on a red sequence, which may indicate the initial establishment of the prominent red sequences of early-type galaxies common in rich, low-redshift clusters (e.g. Coma; \citealt{eis07}). There is also a bright interloper ($K_s = 17.59$) posing as a red sequence member. There are also foreground and background AGN (bright X-ray point sources found with \Chandra) that are entangled within the CMR. The overdensity remains significant, but the aforementioned interlopers urge caution in relying on purely photometric results.

\subsection{3C 068.2}

%\clearpage
\begin{deluxetable}{c c c c c c}
%\tabletypesize{\scriptsize}
\tablewidth{0pt}
\tablecolumns{6}
\tablecaption{3C 068.2 Sources}
\tablehead{
\colhead{Galaxy} & \colhead{$\Delta$RA } & \colhead{$\Delta$DEC} & \colhead{$\Delta$r } & \colhead{$K_s$} & \colhead{$J-K_s$}\\
\colhead{} & \colhead{('')} & \colhead{ ('')} & \colhead{('')} & \colhead{(mag)} & \colhead{(mag)} }
\startdata
1&	0.000&		0.000&		0.000&			18.33$\pm$0.17&	1.60$\pm$0.24	\\
2&	-13.247&	39.826&	41.971&		18.61$\pm$0.19&	1.19$\pm$0.24 \\
3&	4.572&		34.854&	35.153&		17.94$\pm$0.15&	1.93$\pm$0.21 \\
4\tablenotemark{a}&	-7.548&	27.028&	28.063&		16.30$\pm$0.10&	1.99$\pm$0.13 \\
5\tablenotemark{a}&	-12.031&	21.261&	24.429&		17.55$\pm$0.13&	2.09$\pm$0.19 \\
6&	31.568&	20.292&	37.528&		19.26$\pm$0.25&	0.92$\pm$0.32 \\
7&	53.111&	5.645&		53.410&		18.84$\pm$0.21&	0.94$\pm$0.26 \\
8&	-14.288&	0.963&		14.321&		18.68$\pm$0.19&	0.78$\pm$0.23 \\
9&	-46.977&	-6.956&	47.489	&		18.34$\pm$0.17&	0.98$\pm$0.21 \\
10&	-12.658&	-13.683&	18.640&		18.84$\pm$0.21&	1.04$\pm$0.26 \\
11&	12.944&	-43.318&	45.211&		19.28$\pm$0.25&	1.42$\pm$0.35 \\
12&	-1.630&	-20.314&	20.379&		19.42$\pm$0.27&	1.89$\pm$0.43 \\
13&	16.984&	-29.481&	34.023&		20.17$\pm$0.40&	0.59$\pm$0.47 \\
14&	12.224&	4.321&		12.966&		19.98$\pm$0.38&	4.06$\pm$2.14 \\
15&	-0.121&	-8.203&	8.204&			19.41$\pm$0.29&	2.94$\pm$0.80 \\
16&	-6.447&	1.032&		6.529&			19.67$\pm$0.32&	2.28$\pm$0.61 \\
17&	20.055&	-23.083&	30.578&		19.67$\pm$0.32&	2.64$\pm$0.69 \\
18&	-22.220&	-25.015&	33.459&		20.02$\pm$0.38&	3.30$\pm$1.25 \\
19&	-30.541&	-30.224&	42.968&		19.76$\pm$0.31&	0.52$\pm$0.37 \\
20&	28.521&	-41.641&	50.472&		19.99$\pm$0.37&	1.62$\pm$0.55 \\
21&	35.474&	-39.058&	52.763&		20.89$\pm$0.64&	2.10$\pm$1.21 \\
22&	52.244&	-32.845&	61.710&		19.99$\pm$0.36&	0.99$\pm$0.46 \\
23&	19.391&	4.007&		19.801&		19.91$\pm$0.35&	1.50$\pm$0.50 \\
24&	14.409&	-18.372&	23.348&		19.23$\pm$0.26&	1.63$\pm$0.36 \\
25&	-13.351&	-20.148&	24.170&		19.38$\pm$0.27&	1.96$\pm$0.43 \\
26&	28.742&	-4.167&	29.043&		19.28$\pm$0.26&	3.67$\pm$0.90 \\
27&	22.496&	24.888&	33.549&		19.28$\pm$0.26&	2.36$\pm$0.47 \\
28&	0.204&		22.287&	22.288&		18.95$\pm$0.22&	2.48$\pm$0.41 \\
29&	25.898&	-2.416&	26.010	&		19.37$\pm$0.27&	2.15$\pm$0.45 \\
30&	19.655&	23.165&	30.380&		19.58$\pm$0.30&	2.24$\pm$0.53 \\
31&	26.082&	-34.725&	43.430&		18.74$\pm$0.20&	1.93$\pm$0.31 
\enddata
\tablenotetext{a}{X-ray point source identified with \Chandra \citep{wilk13}.}
\tablecomments{Galaxy candidates within $\sim 1'$, which corresponds to $\sim 0.5 \, Mpc$ at this redshift \citep{wri06}, around 3C 068.2 (Galaxy 1) located at RA=2:34:23.8, DEC=+31:34:17. \label{tab:3Ct}}
% Call as Table ~\ref{tab:3Ct}, but does not work if you just compile in PDF Latex.
\end{deluxetable}

\begin{figure*}
\centering
\includegraphics[scale=0.5]{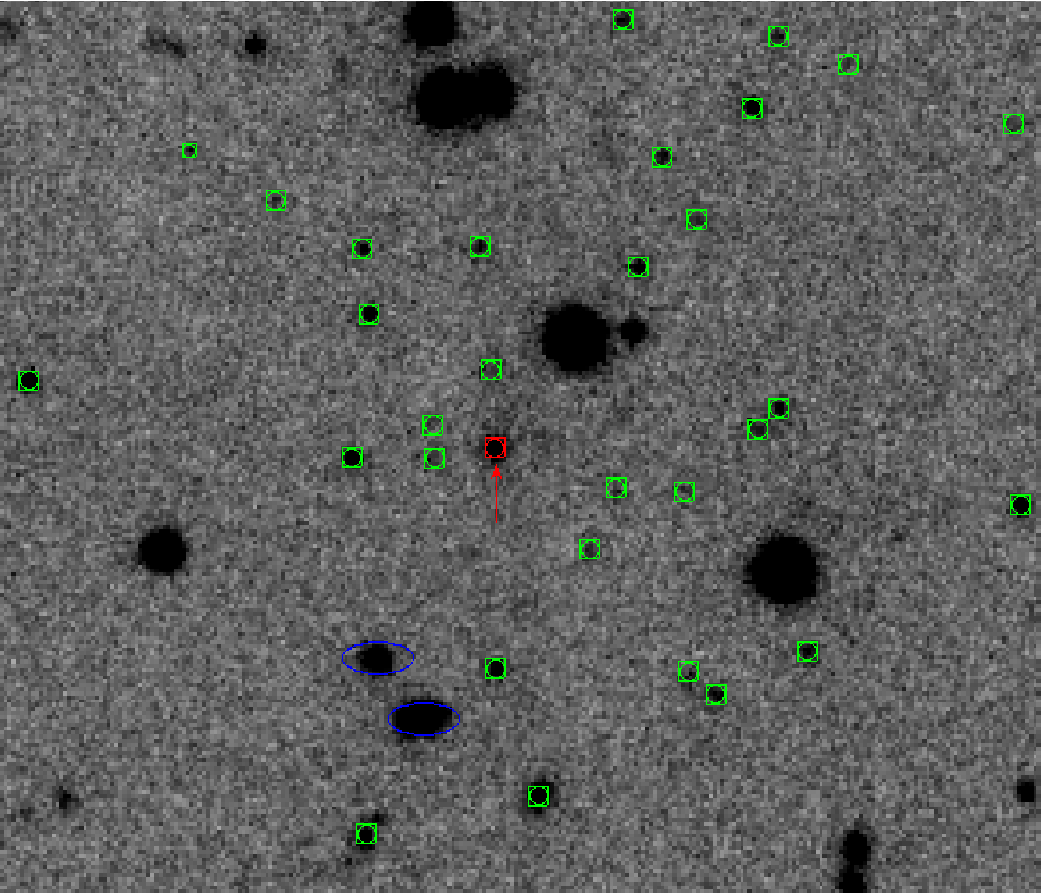}
\caption{\Newfirm view of sources (green boxes) within $\sim 1'$ of 3C 068.2. The radio source is indicated by a red arrow and red square. The blue ellipses are X-ray point sources found with \Chandra \citep{wilk13}.}
\label{fig:3Ci}
\end{figure*}

%\clearpage
\begin{figure*}
\centering
\includegraphics[scale=0.5]{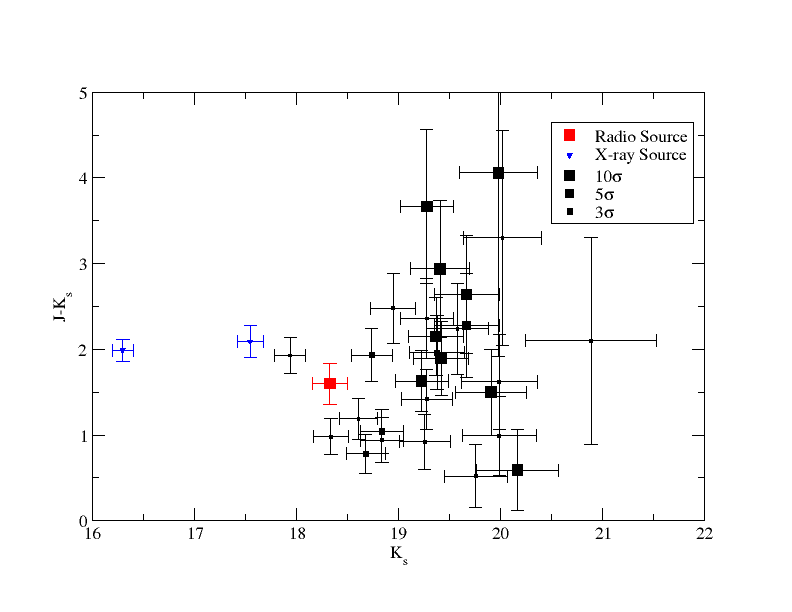}
\caption{A color-magnitude diagram of galaxy sources surrounding 3C 068.2, which is represented by a red square. The radio source is not the brightest cluster galaxy, being located within the blue cloud with a measured star-formation rate of $\sim 390 \, \mathrm{M}_{\odot} \, \mathrm{yr}^{-1}$ \citep{bar12}. The blue, downward triangles are \Chandra point sources \citep{wilk13}, and are the two brightest galaxies on the flat red sequence ($J-K_s \sim 2$). The larger the size of the symbol, the more significant overdensity the galaxy resides in. The largest symbol corresponds to a galaxy within the area corresponding to at least a 10$\sigma$ overdensity. Within this region, $\sim$4 galaxies are expected to be unassociated with the cluster.}
\label{fig:3Cf}
\end{figure*}

%\clearpage

The photometry for the galaxies around 3C 068.2 are in Table~\ref{tab:3Ct}. Within $\sim$1 arcminute of the central radio source (Galaxy 1 in the table), which corresponds to roughly $\sim 0.5$ Mpc in the assumed cosmology \citep{wri06}, 31 extended sources were identified. The selected sources are shown in Fig~\ref{fig:3Ci}.

A $K_s$, $J-K_s$ color-magnitude relation of the sources is in Fig~\ref{fig:3Cf}, displaying a flat red sequence (RS) of $J-K_s \approx 2$. The two bright galaxies on the RS with $K_s =16.30, 17.55$ likely contain AGN, as they were identified as X-ray point sources by \citet{wilk13} with \Chandra. 

It is interesting to note that the central 3C source in this case is neither on the red sequence, nor is it the brightest cluster galaxy. Instead, it is found in the so-called blue cloud, inhabited primarily by star-forming galaxies. \citet{bar12} surveyed a number of 3C sources with \Herschel and integrated the far-infrared (FIR) dust emission to calculate a star formation rate (SFR) of each galaxy. 3C 068.2 has a calculated SFR of $\sim 390$ $\mathrm{M}_{\odot} \, \mathrm{yr}^{-1}$. This exceptionally high SFR cannot be sustained for a Hubble time without exceeding the total mass of the largest known galaxies. This might suggest that this galaxy has been caught in the act of formation prior to its transition from the blue cloud to the red sequence.

\subsection{MS 1426.9+1052}

%\clearpage
\begin{deluxetable}{c c c c c c}
%\tabletypesize{\scriptsize}
\tablewidth{0pt}
\tablecolumns{6}
\tablecaption{MS 1426 Sources}
\tablehead{
\colhead{Galaxy} & \colhead{$\Delta$RA } & \colhead{$\Delta$DEC} & \colhead{$\Delta$r } & \colhead{$K_s$} & \colhead{$J-K_s$}\\
\colhead{} & \colhead{('')} & \colhead{ ('')} & \colhead{('')} & \colhead{(mag)} & \colhead{(mag)} }
\startdata
1	&	0.000	&	0.000	&	0.000	&	17.32	$\pm$	0.18	&	1.65	$\pm$	0.28	\\
2	&	-40.712	&	-35.650	&	54.115	&	19.25	$\pm$	0.22	&	0.87	$\pm$	0.31	\\
3	&	-51.641	&	-29.798	&	59.622	&	18.53	$\pm$	0.19	&	1.53	$\pm$	0.29	\\
4	&	-24.161	&	-28.263	&	37.183	&	18.74	$\pm$	0.20	&	2.27	$\pm$	0.33	\\
5	&	-27.625	&	-29.245	&	40.229	&	17.94	$\pm$	0.18	&	1.81	$\pm$	0.29	\\
6	&	-2.455	&	-22.960	&	23.091	&	19.32	$\pm$	0.23	&	1.86	$\pm$	0.36	\\
7	&	-4.699	&	-14.326	&	15.077	&	19.32	$\pm$	0.23	&	1.19	$\pm$	0.34	\\
8	&	-1.657	&	-8.484	&	8.645	&	17.86	$\pm$	0.18	&	1.89	$\pm$	0.29	\\
9	&	18.830	&	-7.844	&	20.398	&	18.52	$\pm$	0.19	&	1.11	$\pm$	0.29	\\
10	&	10.365	&	-4.190	&	11.180	&	18.31	$\pm$	0.19	&	0.56	$\pm$	0.28	\\
11\tablenotemark{a}	&	-17.572	&	1.746	&	17.658	&	18.88	$\pm$	0.21	&	0.88	$\pm$	0.30	\\
12	&	18.403	&	4.213	&	18.879	&	18.98	$\pm$	0.20	&	1.06	$\pm$	0.30	\\
13	&	-18.602	&	5.316	&	19.347	&	17.98	$\pm$	0.19	&	0.82	$\pm$	0.28	\\
14	&	-6.800	&	7.373	&	10.030	&	18.28	$\pm$	0.19	&	1.74	$\pm$	0.29	\\
15	&	31.851	&	8.302	&	32.915	&	16.84	$\pm$	0.18	&	1.63	$\pm$	0.28	\\
16	&	3.553	&	14.216	&	14.653	&	18.10	$\pm$	0.19	&	1.62	$\pm$	0.29	\\
17	&	16.026	&	21.328	&	26.678	&	18.36	$\pm$	0.19	&	1.24	$\pm$	0.29	\\
18	&	20.009	&	33.412	&	38.945	&	18.95	$\pm$	0.21	&	1.54	$\pm$	0.32	\\
19	&	21.558	&	7.115	&	22.702	&	20.88	$\pm$	0.60	&	-0.34	$\pm$	0.64	\\
20	&	0.422	&	-19.404	&	19.409	&	20.70	$\pm$	0.54	&	0.33	$\pm$	0.61	\\
21	&	-31.632	&	2.164	&	31.706	&	19.91	$\pm$	0.31	&	1.03	$\pm$	0.40	\\
22	&	-3.954	&	67.597	&	67.712	&	18.65	$\pm$	0.20	&	1.01	$\pm$	0.29	\\
23	&	-32.659	&	59.472	&	67.850	&	17.55	$\pm$	0.18	&	2.31	$\pm$	0.29	\\
24	&	-43.508	&	65.673	&	78.777	&	17.94	$\pm$	0.19	&	1.99	$\pm$	0.29	\\
25	&	-36.862	&	67.383	&	76.807	&	17.14	$\pm$	0.18	&	1.71	$\pm$	0.28	\\
26	&	-53.406	&	60.234	&	80.501	&	19.22	$\pm$	0.23	&	1.52	$\pm$	0.34	\\
27	&	-76.388	&	0.576	&	76.390	&	18.82	$\pm$	0.23	&	2.78	$\pm$	0.78	\\
28	&	3.388	&	48.021	&	48.140	&	19.78	$\pm$	0.28	&	0.97	$\pm$	0.37	\\
29	&	-51.012	&	36.668	&	62.823	&	20.21	$\pm$	0.38	&	0.43	$\pm$	0.46	\\
30	&	14.034	&	13.015	&	19.141	&	19.44	$\pm$	0.24	&	2.52	$\pm$	0.54	\\
31	&	12.620	&	34.705	&	36.928	&	19.43	$\pm$	0.25	&	2.62	$\pm$	0.57	\\
32	&	-62.294	&	-16.795	&	64.518	&	19.00	$\pm$	0.21	&	1.97	$\pm$	0.34	\\
33	&	-44.385	&	-11.477	&	45.845	&	19.45	$\pm$	0.27	&	2.11	$\pm$	0.45
\enddata
\tablenotetext{a}{Has a measured photometric redshift of $z=1.265$ using SDSS data \citep{ric09}.}
\tablecomments{Galaxies surrounding Galaxy 1 (RA=14:29:24.1 and DEC=+10:39:15), the radio source MS 1426.9+1052.  \label{tab:MSt}}
\end{deluxetable}

\begin{figure*}
\centering
\includegraphics[scale=0.5]{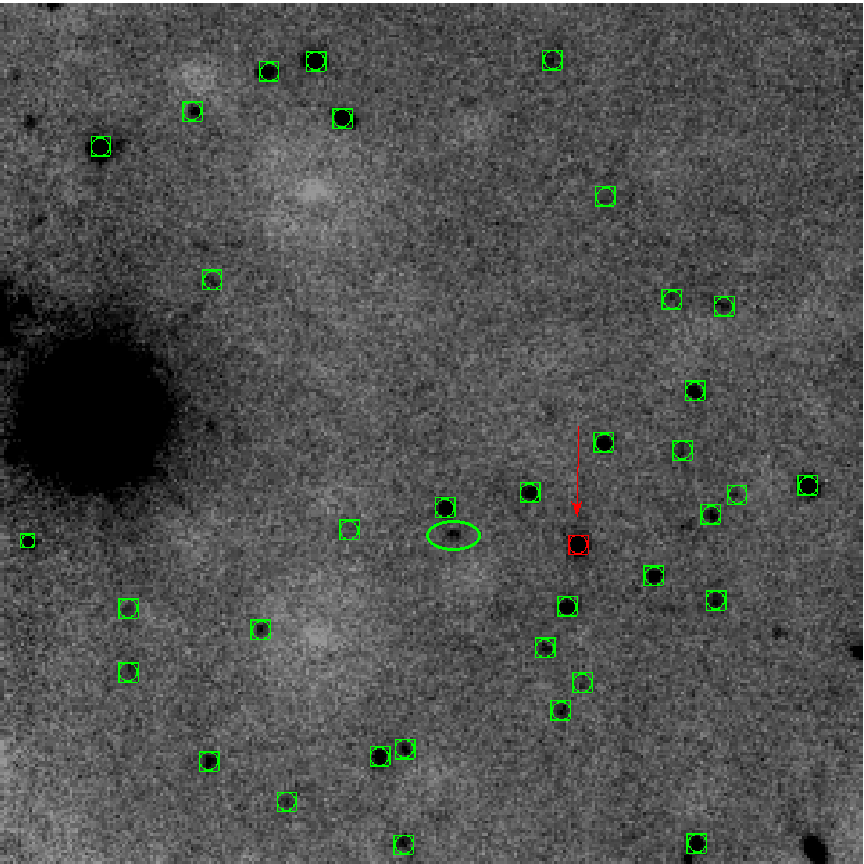}
\caption{$K_s$ image around MS 1426.9+1052 with the \Sqiid camera. The large, saturated object on the left-hand side of the image is a $K_s = 9.17$ mag star in the 2MASS Point-Source Catalog. The radio source is indicated by a red arrow and box, while the green ellipse indicates a galaxy with a photometric redshift of $z=1.265$ \citep{ric09}. The green boxes are the remainder of galaxy sources found within the field. The image size is $120'' \times 120''$.}
\label{fig:MSi}
\end{figure*}

%\clearpage
\begin{figure*}
\centering
\includegraphics[scale=0.5]{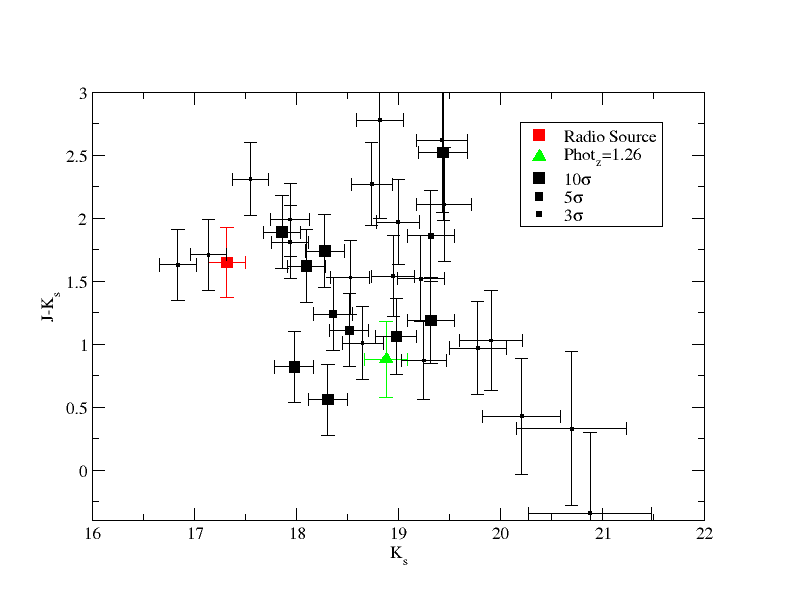}
\caption{The color-magnitude diagram for galaxies surrounding the central radio source (red square). The green upwards triangle is a probable interloping galaxy with a photometric redshift of $z=1.265$ \citep{ric09}, when compared to the redshift of the radio source at $z=1.83$. The largest symbols represent galaxies found within the area corresponding to a 10$\sigma$ overdensity, of which roughly 4 are expected to be field galaxies.}
\label{fig:MSf}
\end{figure*}

MS 1426 is the most distant candidate cluster in this sample, and it contains a large number of potential cluster galaxies, with 34 measured photometrically in Table~\ref{tab:MSt}. The \Sqiid image (Fig~\ref{fig:MSi}) surrounding the radio source was slightly off-center, so the selected galaxies are not evenly distributed from the radio source.  The color-magnitude relation is shown in Fig~\ref{fig:MSf}, with a RS color consistent with the CMRs for CL J0442+0202 (Fig~\ref{fig:CLf}) and surrounding 3C 068.2 (Fig~\ref{fig:3Cf}).

\citet{ric09} computed photometric redshifts of regions surrounding radio-loud sources within the SDSS database. There are two galaxies near MS 1426.9+1052 with photometric redshifts of $z = 1.83$, but were unfortunately off of the edge of the \Sqiid image. However, this provides a further indication that the field hosts a possible structure, in addition to the presence of a red sequence in the color-magnitude relation and the associated galaxy overdensity that was discussed in \secref~\ref{sec:over}.

\section{Discussion} \label{sec:jim}

\begin{figure*}
\centering
\includegraphics[scale=0.5]{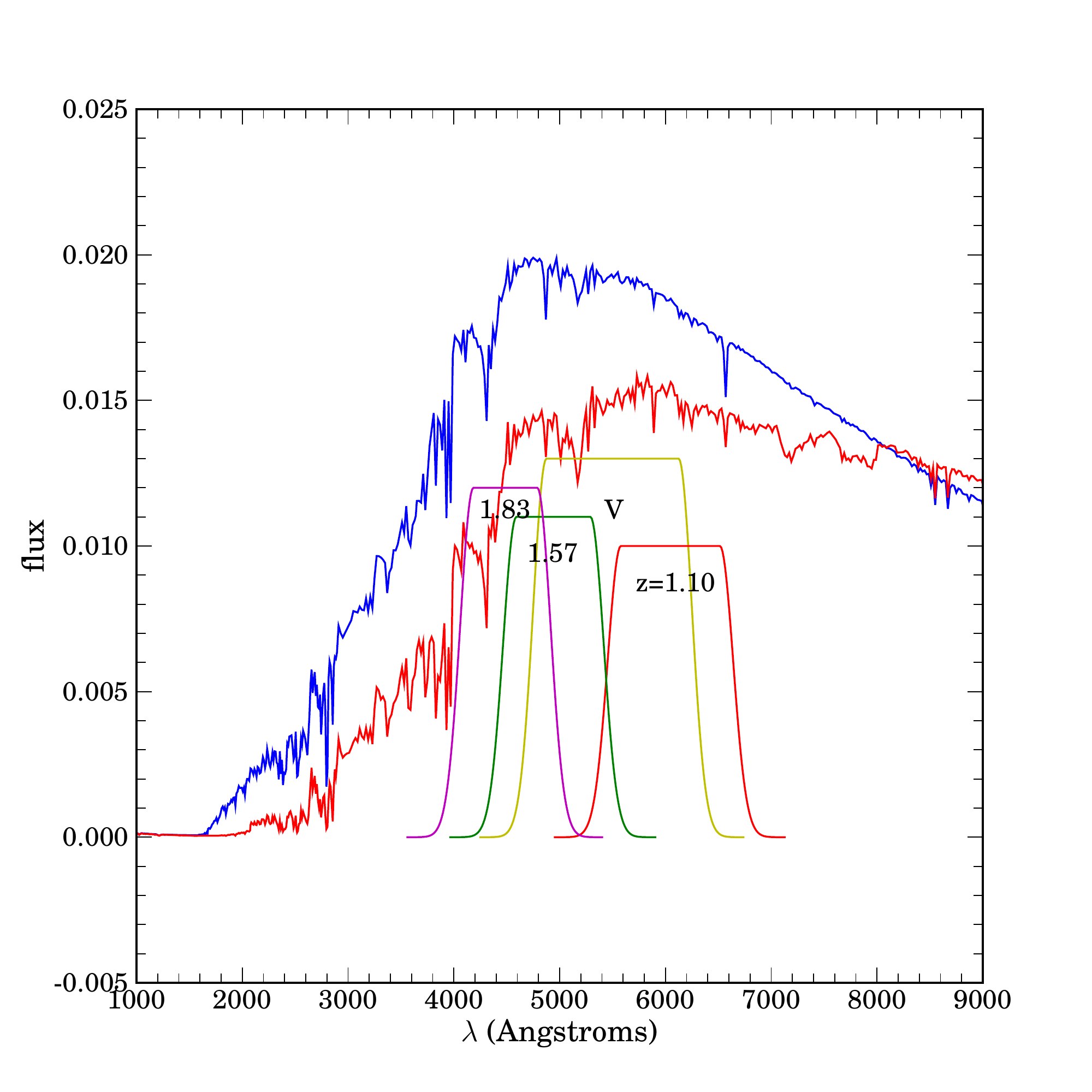}
\caption{Model SED for an old (12 Gyr) galaxy in red, with the blue spectrum illustrating a young galaxy more representative of the expected SED at the high redshift of the cluster candidate galaxies. Overlaying the SEDs are the de-redshifted $J$ filters for each candidate cluster, with $V$ as a reference. At increasing redshifts, the observed filters narrow and measure less flux, so $K$-corrections were applied to the computed $M_{5500}$ AB magnitudes for each source. Flux in arbitrary units.}
\label{fig:Jfil}
\end{figure*}

\begin{figure*}
\centering
\includegraphics[scale=0.5]{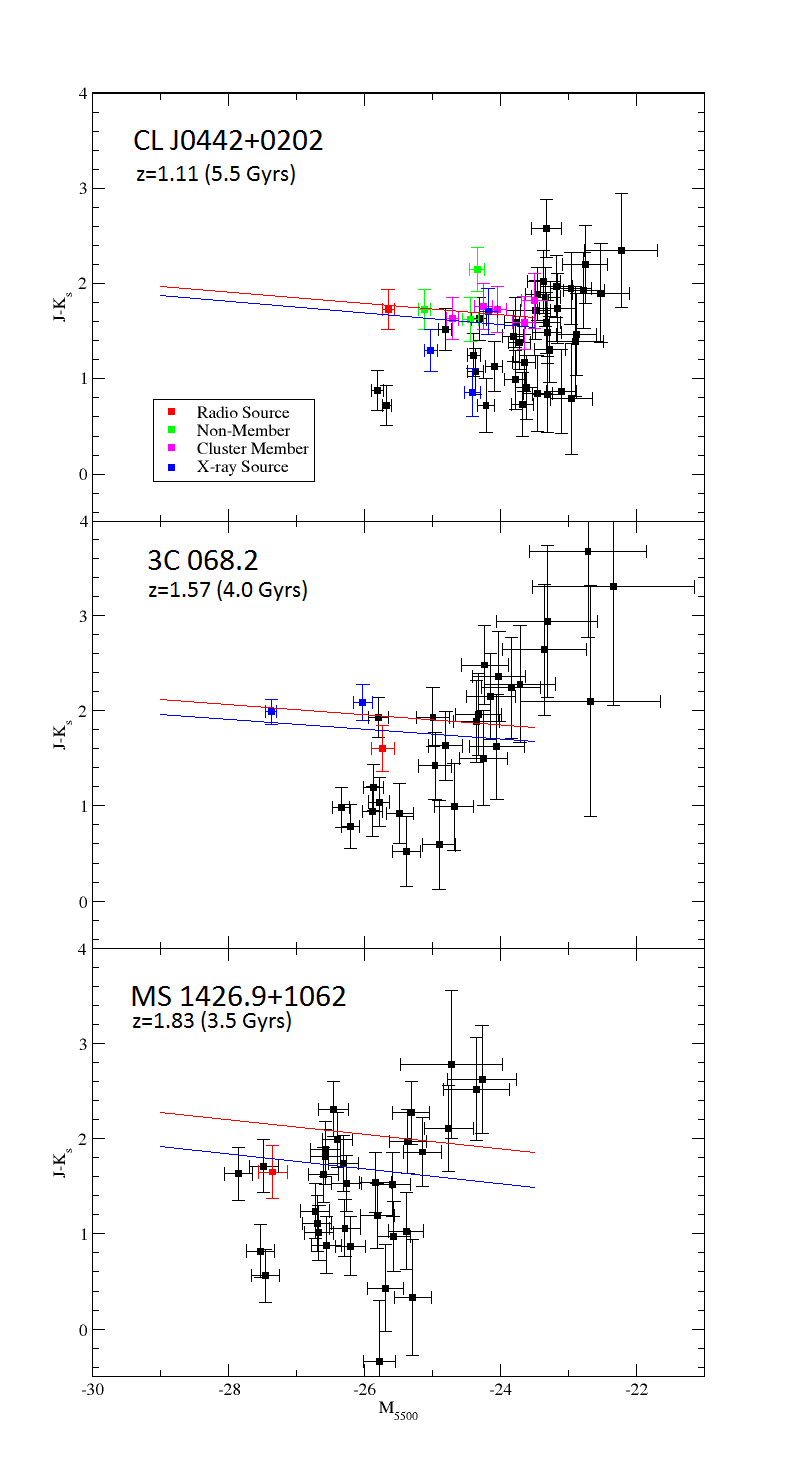}
\caption{Each candidate clusters CMR, with absolute $M_{5500}$ AB magnitudes computed with the assumption that all sources are at the same redshift as the radio-loud source. As in the case of CL J0442+0202, there will be foreground and background interlopers within these CMRs. A $K$-correction has been applied to each cluster candidate based on the SED template age and redshift with respect to the $M_{5500}$ magnitude system (Fig ~\ref{fig:Jfil}). The red CMR line represents the red sequence from the 12 Gyr galaxy models of \citet{scho09a}, redshifted to the respective value. The blue lines are recomputed SED models for the assumed ages of each candidate cluster, which are listed in the parentheses next to the redshift of the radio source.}
\label{fig:scho}
\end{figure*}

\begin{deluxetable*}{c c c c}
%\tabletypesize{\scriptsize}
\tablewidth{0pt}
\tablecolumns{4}
\tablecaption{Blue and Faint Red Fractions of Candidate Cluster Galaxies}
\tablehead{
\colhead{Field} & \colhead{Redshift} & \colhead{$f_b$} & \colhead{$f_{fr}$} 
} 
\startdata
CL J0442+0202 & $z = 1.10$ & 40$\pm14$\% (38$\pm29$\%) & 25$\pm14$\% (27$^{+29}_{-27}$\%)  \\ 
3C 068.2	 & $z=1.57$ & 41$\pm18$\% (27$^{+33}_{-27}$\%) & 26$\pm18$\% (48$\pm33$\%)  \\ 
MS 1426.9+1052 & $z=1.83$ & 49$\pm17$\% (60$\pm33$\%) & 24$\pm17$\% (10$^{+33}_{-10}$\%)    
\enddata
\tablecomments{The blue ($f_b$) and faint red fractions ($f_{fr}$) of galaxies for each cluster candidate. Values in parentheses are the fractions contained within the regions of greatest galaxy overdensity ($>10\sigma$). The blue fraction is calculated by using our $z_f=10$ CMR models for each cluster candidate (blue lines in Fig~\ref{fig:scho}) which represents the expected red sequence at that redshift, and specifying the color cut as 0.2 magnitudes bluer than the expected red sequence, following the example of \citet{but84}. The probability distribution of each galaxy that falls below that color cut is deemed part of the blue fraction. The faint red fraction utilizes the same CMR line, and the color cut is 0.2 mags redder, with the additional constraint that each galaxy must be 1.5 $M_{5500}$ AB magnitudes fainter than the spectroscopic BCG in each cluster. Poissonian errors ($\sqrt{N_{obj}}$) are included.
\label{tab:frac}}
\end{deluxetable*}

Perhaps one of the oldest galaxy correlations is the one between galaxy
absolute luminosity and global color, the color-magnitude relation
(i.e., the CMR; \citealt{bau59,fab73,vis77}). The CMR
has been demonstrated to exist from the near-UV \citep{kav07} to
near-IR wavelengths \citep{cha06}.  A majority of these CMR studies
can be summarized such that (1) the correlation is one of redder galaxy
colors with increasing galaxy luminosity, (2) the correlation exists over a
range of galaxy environments, but is strongest in the densest galaxy
regions \citep{coo08}, and (3) there exist distinct red and blue
components to the CMR, with the red component being composed of early-type
galaxies (ellipticals and S0's, see \citealt{bow92,van98,hog04,lop04,bal04,bel04,ber05,mci05,deluc06,gal06}).

The CMR is a member of a number of fundamental relations that relate the
global properties in early-type galaxies, such as mass and kinematics
\citep{dre87}.  Various formation and evolutionary scenarios are
proposed to explain these fundamental relationships \citep{gav96}.
For example, galaxy interactions can remove the gas used for star formation and
produce a relationship between age and galaxy mass \citep{moo99}. Ram
pressure stripping can also strip galactic gas and halt star formation \citep{tec10}.
However, the comparison between spectroscopic indices and galaxy color
indicates that the CMR is most strongly linked to the chemical enrichment
history of a galaxy, as redder colors map directly into increasing
metallicities \citep{tho05,ber05}.  The slope and
scatter of the CMR can be reproduced if the redshift of galaxy formation is
greater than 5 \citep{bow98} even with a small amount
of mass growth by mergers or bluer colors with recent star formation.

With respect to cluster populations, the CMR has also been used to define
the `red envelope' or red sequence \citep{ell88}, the boundary between old,
quiescent galaxies and star-forming systems.  The small amount of scatter
for the red sequence of the CMR has been used to argue that early-type
galaxies are coeval and have evolved passively since their formation epoch
\citep{mci05,scho09a}.  The blue population (the
`blue cloud') can be compared to the red population by assigning the
fraction of galaxies below 0.2 magnitudes of the red sequence; this fraction ($f_B$)
becomes an indicator of the global star formation history in cluster
populations.  It can also be followed as a function of redshift
(the Butcher-Oemler effect; \citealt{but84}), a key parameter for numerous galaxy formation
and evolution scenarios.

The examination of the CMR in our sample begins with the construction of a
fiducial CMR based on a sample of 1104 ellipticals in the local Universe
\citep{scho09a}.  The midline to that CMR, based on narrow band
colors between 4100\AA\ and 5500\AA, is well described by the 12 Gyr 
old multi-metallicity models constructed in \citet{scho09b}.  These
models adopt the SED spectra from \citet{bru03} for single
metallicity and age populations and convolve each age/metallicity with a
star formation history (initial burst) and chemical enrichment scenario as
outlined in \citet{scho09b}.  The result is a perfect match to
the optical CMR using only old (12 Gyrs) populations of varying metallicity.
Interpretations of the CMR implying the need for younger stars (frosting models, see \citealt{tra00}) 
appear to be the consequence of neglecting the old, metal-poor (i.e., blue)
component of the metallicity distribution \citep{scho09b}. This has been
independently confirmed by \citet{breg06} analyzing the infrared SEDs of
early type galaxies.

The CMR defined by optical colors is not directly transferable to this
project, as the $JK$ observations for this sample do not use redshifted
filter sets as employed by the Rakos \& Schombert program.  Thus, we must
convolve the $JK$ bandpasses to the rest frame model spectra.  The three
cluster redshifts (1.10, 1.57, 1.83) place the center of the $J$ filter
at rest frames 6,000\AA, 4,903\AA, 4,452\AA\, and the $K$ filter at 10,571\AA, 8,775\AA,
7,845\AA, respectively.  This places the observed colors between the
Johnson $B$ and $I$ range, redward of the 4000\AA\ break.  
As the typical elliptical has a spectral intensity peak at 4,600\AA\
(before a sharp drop off at the 4,000\AA\ break and the metal line rich
region between 4,200\AA\ and 4,800\AA), our expectation is that the
observed $J-K$ colors will increase (redden) from $z=$1.10 to 1.57, then
shift blueward at $z=$1.83 as the $J$ filter passes over the 4,600\AA\
bump, as seen in Fig~\ref{fig:Jfil} for both a young and old galaxy.
Indeed, that is what is seen in the mean cluster colors of $J-K_s=1.47\pm0.04$ (CL
J0442+0202), $J-K_s=1.59\pm0.06$ (3C 068.2) and $J-K_s=1.46\pm0.06$ (MS 1426.9+1052), which is the 
 uncertainty-weighted mean of all candidate galaxies.

Comparison to the 12 Gyr old model CMR of \citet{scho09a} is made by redshifting the model spectra to the
redshift of the sample clusters.  The result is shown in
Fig~\ref{fig:scho} (red lines). These CMRs typically lie
redward of the observed colors.  This is, of course, due to the fact that
the model CMR is produced for 12 Gyr populations.  The age of the universe at 
the redshifts of the clusters will be $\sim$5.5, 4 and 3.5 Gyrs, respectively.  

Altering the CMR for age simply requires recalculating the population model by
halting at a younger age.  The only color changes are due to age effects
(shifts in the turnoff point to the blue) as metallicity effects are minor
since single burst models have a rapid enrichment scenario (e.g., the
typical early-type galaxy reaches 90\% final metallicity in 3 Gyrs due to the rapid
onset of galactic winds).  Recalculating the models (and chemical
enrichment) to those cluster ages produces the blue CMR lines in Fig~\ref{fig:scho}.
While the correction due to age is relatively small, the age corrected CMR's
are in better agreement with the cluster data.  The close match between the red
sequence and the model CMR's suggests that these rich cluster ellipticals
are effectively coeval with an early formation epoch. 

The formation redshifts of these cluster candidates are difficult to constrain
within the resolution of our data. For instance, using $z_f = 5$ within
our simulations produces red sequences that are offset from the blue lines
in Fig~\ref{fig:scho} by $\Delta(J-K_s) = -0.08, -0.09, -0.12$ for redshifts of
$z=1.1,1.57,1.83$, respectively. As discussed in \secref~\ref{sec:cand}, the $K'$ photometry of 10 galaxies in CL J0442+0202
from \citet{ste03} show a bias when compared to our $K_s$ photometry of $K' - K_s = 0.32 \pm 0.16$.
For these ten galaxies, this bias could imply a bluer stellar population, and thus a smaller $z_f$.
Within the uncertainties of our data, we cannot formally fit a $z_f$ to the cluster candidates, but the early-type galaxies are
consistent with $z_f \gtrsim 5$.

Constraining the formation age
of the galaxy in high redshift is challenging because of the inverse relation between redshift and cosmic age. For example, \citet{eis08}
found that for a sample of IRAC-selected clusters at redshifts $z > 1$,
$I-[3.6]$ colors favored $z_f >3$, but could also be consistent with $z_f =30$ (see their Fig 19)
with significant scatter in clusters at redshifts above $z>1.5$. Our results are consistent, with at 
least a few galaxies appearing on the red sequence as old populations for their redshift.

Absolute $M_{5500}$ AB magnitudes were computed by convolving the redshifted $J$
filter that each cluster is observed in with $V$. As illustrated in Fig~\ref{fig:Jfil}, the
filters are compressed at higher redshift, which lowers the flux through the filter, so
a $K$-correction of -0.23, -0.50, and -0.75 magnitudes was applied to CL J0442+0202, 
3C 068.2, and MS 1426.9+1052, respectively.

Clear, early evidence of evolution in the stellar populations of cluster
galaxies is the Butcher-Oemler effect \citep{but84}.  Simply
stated, the Butcher-Oemler effect is a sharp increase in the fraction of
blue galaxies in a cluster as a function of redshift.  \citet{rak95} studied 509 galaxies in 17 clusters between
redshifts of 0 and 1, and found a linearly increasing $f_B$ from present-day
value of 5\% to 80\% at a redshift of 1. That study utilized Str{\"o}mgren
photometry to determine a color separation between nearby star-forming galaxies
and ellipticals (anything blueward of $bz-yz=0.20$ or $vz-yz=0.40$), and then used customized filter sets to match those
rest frame colors at the respective redshifts of each cluster. Any source more blue
than this color separation is considered a member of the blue fraction for the cluster.

%J-K=1.3 (rest frame) only works for CL
To calculate the blue fraction using $J,K_s$ filters, we computed the probability
distribution for the color of each galaxy, and the fraction that falls 0.2 mags bluer than the $z_f=10$ CMR models
of the red sequence (blue lines in Fig~\ref{fig:scho}) are considered to be blue galaxies. The 0.2 magnitude offset
is adopted from the example of \citet{but84}, and we assumed Gaussian errors for the probability distribution. This results in
$f_B$ values of 40$\pm14$\%  in CL J0442+0202, 41$\pm18$\%  for 3C 068.2, and 49$\pm17$\%  within MS 1426.9+1052. The $f_B$
are minimally reduced ($\sim 5\%$) if the blue fraction is simply defined as the number of galaxies
that are 0.2 mags bluer than the $z_f=10$ CMR models. Errors are based on an assumed Poisson distribution ($\sqrt{N_{obj}}$), as in \citet{rak95}. 
A summary of the $f_B$ values can be found in Table~\ref{tab:frac}.

The $f_B$ values are less than expected from the rapid increase in $f_B$ from
\citet{rak95} (see their Figure 4).  However, this can be interpreted as a
consequence of the cosmic star formation rate as a function of redshift, in which
galaxies at $z>1$ have nearly constant SFR \citep{mad96}, while the rate steadily
declines from $z=1\rightarrow0$. Updated versions of
the ``Madau diagram'' confirm this trend out to $z\sim3$ \citep{per05}.  Our
values  of the blue fraction also agree with the more intensive spectroscopic GMASS study \citep{cas08}
which finds $f_B$ of 50\% at redshifts of 2. Indeed, \citet{new14} find a cluster at $z=1.8$ that hosts 15/19 spectroscopically
confirmed, quiescent members, further corroborating blue fractions that are
sometimes small at this epoch. These differences could point to large variations
in stages of cluster assembly and evolution in the Universe at these high redshifts.

If the sample is restricted to the regions of maximum overdensity within each cluster (e.g. $>10\sigma$,
 discussed in \secref~\ref{sec:over}), the blue fraction decreases in the two
low redshift targets but increases for MS 1426.9+1052. CL J0442+0202 has a
relatively minor decrease to $f_B=38\pm29$\%, while 3C 068.2 at $z=1.57$ decreases from $f_B=41\pm18\% \rightarrow 27^{+33}_{-27}\% $.
MS 1426.9+1052 at $z=1.83$ increases its blue fraction within the region of largest overdensity 
by 11\% to $f_B=60\pm33$\%. The large variability between candidate clusters at a range of redshifts is an
interesting test of the evolution for cluster galaxies, as the innermost regions
should have had the most dynamical interaction.

We also define the faint red fraction of galaxies ($f_{fr}$) as those galaxies that
are 1.5 $M_{5500}$ AB magnitudes fainter than the spectroscopically confirmed BCG
for each candidate cluster and 0.2 mags redder than the CMR model red sequence line
used to compute the blue fraction, once again using the probability distribution
with assumed Gaussian errors for each galaxy. CL J0442+0202 has a $f_{fr}=25\pm14$\% for the entire sample
of galaxies measured, and $f_{fr}=27^{+29}_{-27}$\% within the region of $>10\sigma$ overdensity.
At the redshift of $z=1.57$, 3C 068.2 has a $f_{fr}=26\pm18$\% overall, but a 22\% increase
within the central overdensity to $f_{fr}=48\pm33$\%, which corroborates the results
from its $f_b$ measurement. Our highest redshift candidate MS 1426.9+1052 has a 
comparatively low red fraction within the overdense region of a mere $f_{fr}=10^{+33}_{-10}$\%, 
with $f_{fr}=24\pm17$\% for the entire sample. These faint, red galaxies are presumably reactants
in future dry mergers that would be hierarchically accreted to form galaxies
on the red sequence. The blue and faint red fractions are listed in Table~\ref{tab:frac}.

Simple scenarios to build
up the red sequence via the quenching and merging of blue galaxies, followed by dry
mergers have been proposed by \citet{fab07}. These
 would require a significant population of interacting blue disk galaxies
to transform relatively weak CMRs at high redshift into the developed red sequence features
seen in rich, nearby clusters (see Figs 3, 4 in \citealt{eis07}). \citet{pap12} conclude
that a series of dry mergers is the dominant mechanism in the growth of quiescent galaxies
based on an in-depth analysis of a cluster at $z=1.62$. The processes that shape the growth
of even the most massive galaxies do not seem to be resolved. Some recent studies conclude that
ongoing mergers after $z\sim1.5$ are necessary to match the present-day BCGs \citep{fer12,fas14},
while \citet{fri14} suggest that major mergers are of minimal importance.

The extent to which there is a well defined red sequence CMR at high redshift impacts
our models of galaxy formation.  CMRs out to redshifts of 1 are
well established (see \citealt{mei09} and references therein) as well as in some 
protoclusters at redshifts of up to $z=4$ \citep{ste05,ove08}.
The stellar populations of galaxies on
the red sequence at $z=1.5$, which match passively evolving color
evolution models, is consistent with the massive, bright end
of the CMR being in place by a few Gyrs after the dawn of time.
This suggests that at least some massive galaxies experience the 
epoch of primary star formation at quite high redshift. Our results 
are consistent with the findings of \citet{eis08} (see their Figs 17,19).

This is in agreement with the observation that protocluster galaxies are,
in general, older than field galaxies of the same redshift \citep{pet07},
as if high density environments accelerate galaxy color and
morphological evolution.  Larger intrinsic scatter in the CMR
with redshift has been interpreted as due to age effects \citep{mei09};
however, it becomes increasingly difficult to separate passive age changes
in ellipticals with increasing star formation in the blue cloud.  The
separation between the red sequence and the blue cloud becomes problematic
inside of 2 Gyrs from the onset of galaxy formation.

The red sequence is nominally identified with ellipticals in the cluster
core region, for they have the reddest colors out to high redshifts \citep{mei09}.
This is supported by the fact that the slope and zeropoint of
the CMR is unchanged (aside from age corrections) between our sample (at a mean
redshift of 1.5) and the present-day.  The galaxies producing an increasing blue
fraction with redshift are perhaps the cluster proto-S0 population, for they
have the largest color scatter in present-day clusters and would match the
``quenching" scenario, where infalling galaxies interact with the
cluster environment to halt star formation \citep{rak95,fab07}.

While the results presented herein are similar to other studies based on
spectroscopic and space-imaging morphology, we achieve many of the same
goals with respect to the star formation history of cluster galaxies, but with a
much smaller amount of telescope time.  In addition, the identification of
the high redshift clusters themselves, as preliminary observations before
spectroscopy on large aperture telescopes and limited space-imaging time,
argues for an extensive ground-based near-IR survey of the environment
around radio sources.  The analysis of the photometric CMR, combined with
accurate stellar population models, provides a different window into the
star formation history of cluster galaxies.

\section{Summary and Future Work} \label{sec:sum}

We have presented three candidate galaxy clusters spanning the redshift regime of $z=1.10 - 1.83$ identified as near-infrared galaxy overdensities surrounding radio-loud sources. From deep NIR imaging, color magnitude diagrams were constructed and compared to the galaxy models of \citet{scho09a}. We find that the ellipticals on the red sequence within each cluster candidate were all formed at a relatively old age. The data are consistent with $z_f\gtrsim 5$ in our models, but we lack the resolution to formally fit a formation redshift. These apparently old, massive red sequence galaxies, while clearly present, are a minority of the cluster population. There is a slowly increasing blue fraction of galaxies within each cluster candidate as a function of redshift, from $f_B = 40\% $ at $z=1.11$ up to $f_B = 48\% $ at $z=1.83$.

The candidate clusters are prime targets for further study. Use of the \emph{Hubble Space Telescope's} superior imaging capabilities, compared to ground-based NIR images, could provide galaxy morphology and mass estimates \citep{cas13}, along with measurements of individual SFRs \citep{lee14b}, even at these redshifts. The two candidates with larger redshifts, 3C 068.2 and MS 1426.9+1052, would benefit from multi-object spectroscopy, primarily to establish cluster membership, while secondarily providing insight into the types of galaxies inhabiting proto-clusters (e.g. emission line signatures), as \citet{ste03} did with CL J0442+0202. 

In the X-ray, \Chandra or \Xmm could potentially be used to confirm these objects as clusters via their ICM emission, as well as providing a mass-estimate for the system. Similarly, one could use CARMA to measure the SZ signal, which will accomplish similar goals as the aforementioned X-ray observations, but with a potentially more modest investment of telescope time.

Most importantly, a confirmation of the presence of a physical structure with spectroscopic redshifts in any of these potential clusters will add to the small but growing number of high-redshift clusters. The number density as a function of redshift and an estimate of the structures mass can provide constraints on cosmological parameters and offer comparisons to simulations of large-scale structure. The leverage to constrain models improves with increasing mass and redshift. Further investigations into the properties of the constituent galaxies (morphology, colors, spectral signatures) will be facilitated by first establishing a $quasi-$virialized structure.   

\acknowledgements

We would like to thank the anonymous referee for helpful comments. The work of SSM is supported in part by NASA ADAP grant NNX11AF89G. This research has made use of the NASA/IPAC Extragalactic Database (NED) which is operated by the Jet Propulsion Laboratory, California Institute of Technology, under contract with the National Aeronautics and Space Administration. This publication makes use of data products from the Two Micron All Sky Survey, which is a joint project of the University of Massachusetts and the Infrared Processing and Analysis Center/California Institute of Technology, funded by the National Aeronautics and Space Administration and the National Science Foundation.

\bibliography{protos}
\bibliographystyle{apj}

%EXAMPLE: [protos.bib] file, style file is apj.bst

%CALL AS: 
%	PDFLATEX PROTOS_V3.TEX
%	BIBTEX PROTOS_V3.AUX
%	PDFLATEX PROTOS_V3.TEX
%	PDFLATEX PROTOS_V3.TEX

%@ARTICLE{hal98, ****WHERE hal98 IS NOW THE REFERENCE TO THIS SOURCE***
%   author = {{Hall}, P.~B. and {Green}, R.~F.},
%    title = "{An Optical/Near-Infrared Study of Radio-loud Quasar Environments. II. Imaging Results}",
%  journal = {\apj},
%   eprint = {astro-ph/9806151},
% keywords = {GALAXIES: CLUSTERS: GENERAL, GALAXIES: PHOTOMETRY, GALAXIES: QUASARS: GENERAL, Galaxies: Clusters: General, Galaxies: Photometry, Galaxies: %Quasars: General},
%     year = 1998,
%    month = nov,
%   volume = 507,
%    pages = {558-584},
%      doi = {10.1086/306349},
%   adsurl = {http://adsabs.harvard.edu/abs/1998ApJ...507..558H},
%  adsnote = {Provided by the SAO/NASA Astrophysics Data System}
%}

\end{document}